\documentclass[aps,graphicx,twocolumn]{revtex4}
\usepackage{amsmath}
\usepackage{amscd}
\usepackage{graphicx}
\usepackage{multirow}
\usepackage{subfigure}
\begin{document}

\title{Hyperentanglement concentration for time-bin and polarization hyperentangled photons}
\author{ Xi-Han Li$^{1,2}$\footnote{
Email address: xihanlicqu@gmail.com}, Shohini Ghose$^{2,3}$}
\address{$^1$ Department of Physics, Chongqing University,
Chongqing, China, 401331 \\$^2$Department of Physics and Computer Science, Wilfrid Laurier University, Waterloo, Canada, N2L 3C5\\
$^3$ Institute for Quantum Computing, University of Waterloo, Canada, N2L 3G1}

\date{\today }
\begin{abstract}
We present two hyperentanglement concentration schemes for  two-photon states that are partially entangled in the polarization and time-bin degrees of freedom. The first scheme distills a maximally hyperentangled state from two identical less-entangled states with unknown parameters via the Schmidt projection method. The other scheme can be used to concentrate an initial state with known parameters, and requires only one copy of the initial state for the concentration process. Both these two protocols can be generalized to concentrate $N$-photon hyperentangled Greenberger-Horne-Zeilinger states that are simultaneously entangled in the polarization and time-bin degrees of freedom. Our schemes require only linear optics and are feasible with current technology. Using the time-bin degree of freedom rather than the spatial mode degree of freedom can provide savings in quantum resources, which makes our schemes practical and useful for long-distance quantum communication.
\end{abstract}
\maketitle
\section{Introduction}
Entanglement is a unique quantum mechanical phenomenon that is a crucial resource for quantum information processing. It has been widely used in quantum communication and quantum computation protocols over the past decades \cite{book}. Entangled photon systems can serve as a quantum channel in many long-distance quantum communication schemes such as quantum key distribution \cite{qkd1,qkd2}, dense coding \cite{dense1,dense2},  teleportation \cite{tele}, quantum secret sharing \cite{qss1,qss2,qss3}, and quantum secure direct communication \cite{qsdc1,qsdc2,qsdc3}. Single photons are interesting candidates for quantum
communication due to their manipulability and high-speed transmission, and because they have several degrees of freedom (DOFs) to carry quantum information. This also allows the possibility of entanglement in a single or multiple degrees of freedom. So far, photons entangled in polarization, spatial modes, time-bin, frequency, and orbital angular momentum have all been successfully generated in experiments. Moreover, hyperentanglement in which photons are simultaneously entangled in more than one DOF has also been demonstrated \cite{type11,type12,preparation1,preparation2,preparation3,preparation4,preparation5,preparation6,preparation7,preparation8}.


Hyperentangled states can be used to beat the channel capacity limit of superdense coding with linear optics \cite{dense1,qkd}, construct hyper-parallel photonic quantum computing \cite{computation,hyper4} which can reduces the operation time and the resources consumed in quantum information processing, achieve the high-capacity quantum communication with the complete  teleportation and entanglement swapping in two DOFs \cite{hyper5, hyper6}. They can also help to design deterministic entanglement purification protocols \cite{depp,odepp,odepp1,omdepp} which work in a deterministic way, not a probabilistic one, far different from conventional entanglement purification protocols \cite{puri pdc,puri1,puri2}. They have been used to assist the complete Bell-state analysis \cite{depp,bsa1,bsa2,bsa3,bsa4,gsa}.

However, entangled states will inevitably interact with the environment during  transmission and storage. This
degrades the fidelity and entanglement of the quantum states, which subsequently reduces the fidelity and security of quantum communication schemes. One solution proposed to preserve the fidelity of entangled channels is entanglement concentration. This method can be used to distill maximally entangled states from an ensemble of less-entangled pure states~\cite{concen1}.
Many interesting entanglement concentration schemes considering different physical systems, different entangled states and exploiting different components have been proposed and discussed  \cite{concen swap1,concen swap2,concen pbs1, concen pbs2,concen sheng1,concen sheng4,concen deng}.

Recently, the distillation of hyperentangled
states has attracted much attention since hyperentanglement has increasing applications in quantum information processing.
In 2013, Ren, Du, and Deng \cite{hc1} presented the parameter-splitting method, a very efficient way for entanglement concentration with linear optics, and they gave the first hyperentanglement concentration protocol  for two-photon four-qubit systems, which was extended to multipartite entanglement subsequently \cite{hc5}. Subsequently, Ren and Deng proposed the first hyperentanglement purification protocol and an efficient hyper-ECP assisted by diamond NV centers inside photonic crystal cavities \cite{hc2}. In 2014, Ren, Du, and Deng gave a two-step hyperentanglement purification protocol for polarization-spatial hyperentangled states with the quantum-state-joining method. It has a higher efficiency \cite{hp}.
Recently, Ren \emph{et al } proposed a general hyperentanglement concentration method  for photon systems assisted by quantum-dot spins inside optical microcavities \cite{hc3}. In 2013, one of us proposed two hyperconcentration schemes with known and unknown parameters, respectively \cite{hc4}. Hyperconcentration based on projection measurements was also proposed \cite{hcc}.

All hyperentanglement concentration schemes so far have dealt with a state which is entangled in the polarization and spatial mode DOFs. Here we focus on hyperentanglement concentrations of states entangled in the polarization and time-bin degrees of freedom.  The polarization is the most popular DOF of the photon due to the ease with which it can be manipulated with current technology. The spatial mode is also easy to manipulate and measure with linear optical elements. However, each photon requires two paths during the transmission when we choose the spatial mode to carry information, which can be a significant issues in long-distance multi-photon communication.
The time-bin DOF is also a simple, conventional classical DOF. Two different times of arrival can be used to encode the logical 0 and 1.  The time-bin states can be simply discriminated by the time of arrival. On the other hand, the manipulation of the time-bin DOF is not easy. The Hadamard operation and measurement of the time-bin state in the diagonal basis $\vert \pm\rangle=1/\sqrt{2}(\vert 0\rangle \pm \vert 1\rangle)$ are difficult.

In this paper we show how to manipulate the time-bin and polarization DOFs for hyperconcentration of two-photon entanglement.
Our first scheme uses two less-entangled pairs with unknown parameters to concentrate hyperentanglement via the Schmidt projection method. The second scheme we propose, which only uses one copy of the less-entangled state with known parameters, borrows some ideas from the parameters splitting method \cite{hc1} . Both these two schemes can be generalized to concentrate $N$-photon hyperentangled GHZ states, and the success probability remains unchanged with the growth of the number of photons. Moreover, our schemes do not require nonlinear interactions that are difficult to implement with current technology. The time-bin entanglement is a stable and useful DOF \cite{time} and does not require two paths per photon compared with the spatial modes. Our proposed schemes are thus practical and useful for  long-distance quantum communication based on hyperentanglement.

\section{Hyperentanglement concentration with unknown parameters}
\begin{figure}[!h]
\centering
\includegraphics*[width=3in]{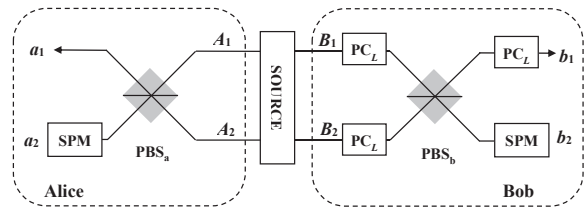}
\caption{Schematic diagram of our scheme for concentration of a hyperentangled state with unknown parameters. Two identical less-entangled state $\vert \phi\rangle_{A_1B_1}$ and $\vert \phi\rangle_{A_2B_2}$ that are originally prepared by the source are shared by two remote parties Alice and Bob. The two parties change the state of $A_2$ and $B_2$ to $\vert \phi'\rangle_{A_2B_2}$ before the concentration. The operations were omitted in this figure. The PBS$_i$ ($i=a,b$) represents a polarizing beam splitter which transmits the horizontal polarization state $\vert H\rangle$ and reflects the vertical polarization state $\vert V\rangle$. PC$_L$ (PC$_S$) is a Pockel cell which effects a bit flip operation when the $L(S)$ component is present. SPM denotes a single-photon measurement which is performed on the second photon of each party. With this device, Alice and Bob implement a parity check of the polarization and the time-bin DOFs, respectively. } \label{c1}
\end{figure}

Suppose the initial two-photon partially hyperentangled state which is entangled in both the polarization and time-bin DOFs can be written as
\begin{eqnarray}
  \vert \phi\rangle_{AB}=
  &&(\alpha \vert HH\rangle +\beta\vert VV\rangle)\nonumber\\
  &&\otimes(\delta\vert SS\rangle+\eta\vert LL\rangle)]_{AB}.
\end{eqnarray}
Here $\vert H\rangle$ and $\vert V\rangle$ represent the horizontal and the vertical polarization states of photons, respectively. $S$ and $L$ denote the two different time-bins, the early ($S$) and the late ($L$). The time interval between the two time-bins is $\Delta t$. The subscript $A$ and $B$ signify the photons held by  two distant parties Alice and Bob, respectively. The four parameters $\alpha$, $\beta$, $\delta$ and $\eta$ are unknown to the two parties and they satisfy the normalization condition $\vert \alpha\vert ^2+\vert \beta\vert ^2=\vert \delta\vert^2+\vert \eta\vert^2=1$. In order to distill the maximally hyperentangled state from the partially entangled ones, two identical original states are required, $\vert \phi\rangle_{A_1B_1}$ and $\vert \phi\rangle_{A_2B_2}$. First, the two parties flip the polarization and time-bin states of $A_2$ and $B_2$, respectively. Then the state changes to
\begin{eqnarray}
  \vert \phi'\rangle_{A_2B_2}=
  &&(\alpha \vert VV\rangle +\beta\vert HH\rangle)\nonumber\\
  &&\otimes(\delta\vert LL\rangle+\eta\vert SS\rangle)]_{A_2B_2}.
\end{eqnarray}
The bit-flip operation of polarization state can be realized by the half wave plate (HWP), while the flip of time-bin state can be completed by the active switches \cite{sw}.
The whole state of the four photons can be written as
\begin{eqnarray}
  &&\vert \Phi_0\rangle_{A_1B_1A_2B_2}\nonumber\\&=&\vert \phi\rangle_{A_1B_1}\otimes \vert \phi\rangle_{A_2B_2}\nonumber\\
 &=&[\alpha^2\vert HHVV\rangle+\beta^2\vert VVHH\rangle\nonumber\\&&+\alpha\beta(\vert HHHH\rangle+\vert VVVV\rangle)]_{A_1B_1A_2B_2}\nonumber\\
 &&[\delta^2\vert SSLL\rangle+\eta^2\vert LLSS\rangle\nonumber\\&&+\delta\eta(\vert SSSS\rangle+\vert LLLL\rangle)]_{A_1B_1A_2B_2}.
\end{eqnarray}

The schematic of our hyperentanglement concentration scheme is shown in Fig. \ref{c1}.
Alice's two photons are incident on a polarizing beam splitter (PBS) PBS$_a$, which is used to perform a polarization parity check on these two photons. The PBS transmits the horizontal states $\vert H\rangle$ and reflects the vertical ones $\vert V\rangle$. If the two photons have the same polarization state, i.e., the even-parity state, there is one and only one photon exiting from each output port of the PBS. Otherwise, two photons exit the same output port when they are in the odd-parity state. Since we cannot distinguish these two photons after the PBS, we use the spatial modes $a_1$ and $a_2$ to denote them. By postselecting the even-parity case the corresponding state is
\begin{eqnarray}
  &&\vert \Phi_1\rangle_{a_1B_1a_2B_2}\nonumber\\
 &=&[\alpha\beta(\vert HHHH\rangle+\vert VVVV\rangle)]_{a_1B_1a_2B_2}\nonumber\\
 &&[\delta^2\vert SSLL\rangle+\eta^2\vert LLSS\rangle\nonumber\\&&+\delta\eta(\vert SSSS\rangle+\vert LLLL\rangle)]_{a_1B_1a_2B_2}.
\end{eqnarray}

Before sending his two photons into a PBS$_b$, Bob uses two Pockel cells (PC) \cite{PC} to flip the polarizations of particles $B_1$ and $B_2$  at a specific time. The PC$_L$ (PC$_S)$ is activated only when the $L$ $(S)$ component is present. Then the state changes to
\begin{eqnarray}
 && \vert \Phi_2\rangle_{a_1B_1a_2B_2}\nonumber\\
 &=&\alpha\beta[\delta^2(\vert H^SH^SH^LV^L\rangle+\vert V^SV^SV^LH^L\rangle)\nonumber\\&&+\eta^2(\vert H^LV^LH^SH^S\rangle+\vert V^LH^LV^SV^S\rangle)\nonumber\\&&+\delta\eta(\vert H^SH^SH^SH^S\rangle+\vert V^SV^SV^SV^S \rangle\nonumber\\&&+\vert H^LV^LH^LV^L\rangle+\vert V^LH^LV^LH^L\rangle)]_{a_1B_1a_2B_2}.
\end{eqnarray}
Here $\vert H^S\rangle$ indicates that the polarization state is $\vert H
\rangle$ while the time-bin state is $\vert S\rangle$. Then PBS$_b$ is utilized to compare the parity of the polarization states of $B_1$ and $B_2$ and the even-parity case is postselected. Actually, due the effect of PCs, Bob's device in effect compares the parity of the time-bin state of $B_1$ and $B_2$. With the effect of another PC$_L$ on path $b_1$, the state of the four photons finally becomes
\begin{eqnarray}
 && \vert \Phi_3\rangle_{a_1b_1a_2b_2}\nonumber\\&=&\alpha\beta\delta\eta(\vert H^SH^SH^SH^S\rangle+\vert V^SV^SV^SV^S\rangle\nonumber\\&&+\vert H^LH^LH^LV^L\rangle+\vert V^LV^LV^LH^L\rangle)_{a_1b_1a_2b_2}.
\end{eqnarray}
The two parties can obtain this state with probability $4|\alpha\beta\delta\eta|^2$.

The last step is to get one of the four maximal hyperentangled states $\vert \Psi_{\pm\pm}\rangle_{AB}$ from $\vert \Phi_3\rangle_{a_1b_1a_2b_2}$ by measuring photons on paths $a_2$ and $b_2$ appropriately.
\begin{eqnarray}
  \vert \Psi_{\pm\pm}\rangle_{AB}=
  &&\frac{1}{\sqrt{2}}(\vert HH\rangle \pm\vert VV\rangle)\nonumber\\
  &&\otimes\frac{1}{\sqrt{2}}(\vert SS\rangle\pm\vert LL\rangle)]_{AB}.
\end{eqnarray}

\begin{figure}[!h]
\centering
\includegraphics*[width=2.5in]{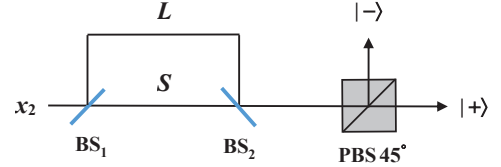}
\caption{(Color online) Schematic diagram of the single-photon measurement setup which consists of only passive linear optics. Here $x_2$ can be $a_2$ $(b_2)$ for Alice (Bob). BS denotes the 50:50 beam splitter. The PBS oriented at $45^o$ transmits the $\vert +\rangle$ polarization states and reflects the $\vert -\rangle$ ones. It is used to measure the polarization state in the diagonal basis.  } \label{s1}
\end{figure}

The first single-photon measurement (SPM) setup consists of only linear optical elements as shown in Fig.\ref{s1}. Two beam splitters (BSs) are used to build an unbalanced  interferometer (UI). The length difference between the two arms is set exactly to $c\Delta t$, where $c$ is the speed of the photons. The effect of the UI can be described by
\begin{eqnarray}
  \vert X^L\rangle \rightarrow \frac{1}{\sqrt{2}}(\vert X^{LS}\rangle+\vert X^{LL}\rangle),\nonumber\\
  \vert X^S\rangle \rightarrow \frac{1}{\sqrt{2}}(\vert X^{SS}\rangle+\vert X^{SL}\rangle).
\end{eqnarray}
Here $X$ denotes $H$ or $V$, and $X^{ij}$ $(i,j,=L,S)$ means the time-bin $i$ pass through the path $j$ of the UI.  After the UI, the state can be written as
\begin{eqnarray}
  &&\vert H^SH^S\rangle_{a_1b_1}\otimes(\vert H^{SS}\rangle+\vert H^{SL}\rangle)_{a_2}\otimes(\vert H^{SS}\rangle+\vert H^{SL}\rangle)_{b_2}\nonumber\\
  &&+\vert V^SV^S\rangle_{a_1b_1}\otimes(\vert V^{SS}\rangle+\vert V^{SL}\rangle)_{a_2}\otimes(\vert V^{SS}\rangle+\vert V^{SL}\rangle)_{b_2}\nonumber\\
  &&+\vert H^LH^L\rangle_{a_1b_1}\otimes(\vert H^{LS}\rangle+\vert H^{LL}\rangle)_{a_2}\otimes(\vert V^{LS}\rangle+\vert V^{LL}\rangle)_{b_2}\nonumber\\
  &&+\vert V^LV^L\rangle_{a_1b_1}\otimes(\vert V^{LS}\rangle+\vert V^{LL}\rangle)_{a_2}\otimes(\vert H^{LS}\rangle+\vert H^{LL}\rangle)_{b_2}.\nonumber\\
\end{eqnarray}

The $LS$ and $SL$ components will arrive at the same time. Therefore, there are three time slots for each particle $a_2$ and $b_2$ to be detected; the middle slot $LS (SL)$, an early slot $SS$ and a late slot $LL$.  The PBS oriented at $45^\circ$ reflects the $\vert -\rangle$ states and transmits the $\vert +\rangle$ ones, where $\vert \pm\rangle=\frac{1}{\sqrt{2}}(\vert H\rangle\pm\vert V\rangle)$. We thus find that only when the two photons are both detected in the middle time slot $LS$ or $SL$ will the collapsed state of $a_1$ and $b_1$ be maximally hyperentangled. The probability of this outcome is  $1/4$. The relation between measurement results of $a_2b_2$ and the final state of $a_1b_1$ is shown in Table I. Otherwise, the state of photons $a_1$ and $b_1$ is only entangled in the polarization DOF. Taking  the probabilities of the two measurement steps into consideration, the total success probability of obtaining a maximally hyperentanged state is $P_0=\alpha^2\beta^2\delta^2\eta^2$, which is a quarter of that of the hyperconcentration scheme for polarization and spatial mode hyperentanglement \cite{hc1}. This is because it is much more difficult to manipulate the temporal DOF.

\begin{table}[h]
\caption{The relation between measurement results of $a_2b_2$ in the middle time slot and the final state of $a_1b_1$. }\label{tab1}
\begin{tabular}{c|ccl }
  \hline
   $M_{a_2b_2}$  &$$& $\vert \Psi\rangle_{a_1b_1}$ \\ \hline \hline
  $\vert +\rangle_{a_2}\vert +\rangle_{b_2}$ &$$& $\frac{1}{2}(\vert HH\rangle+\vert VV\rangle)\otimes(\vert SS\rangle+\vert LL\rangle)$ &   \\ \hline
  $\vert +\rangle_{a_2}\vert -\rangle_{b_2}$ &$$& $\frac{1}{2}(\vert HH\rangle-\vert VV\rangle)\otimes(\vert SS\rangle-\vert LL\rangle)$ &   \\ \hline
  $\vert -\rangle_{a_2}\vert +\rangle_{b_2}$ &$$& $\frac{1}{2}(\vert HH\rangle-\vert VV\rangle)\otimes(\vert SS\rangle+\vert LL\rangle)$ &   \\ \hline
  $\vert -\rangle_{a_2}\vert -\rangle_{b_2}$ &$$& $\frac{1}{2}(\vert HH\rangle+\vert VV\rangle)\otimes(\vert SS\rangle-\vert LL\rangle)$ &   \\
  \hline
\end{tabular}
\end{table}

\begin{figure}[!h]
\centering
\includegraphics*[width=3in]{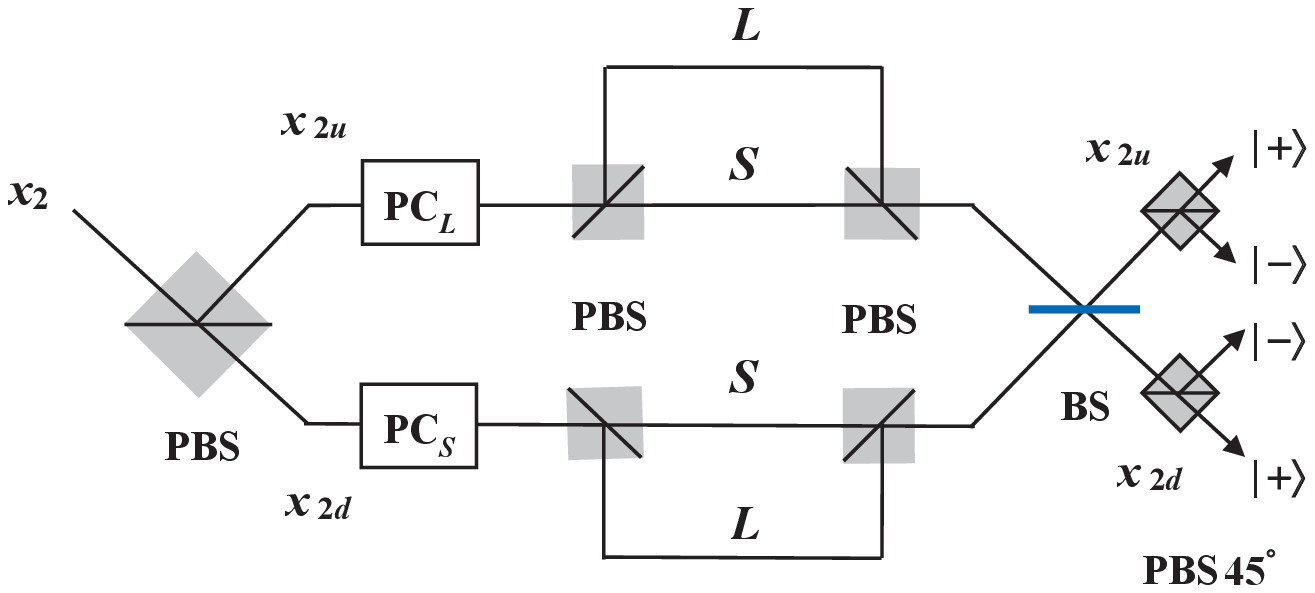}
\caption{(Color online) Schematic diagram of the improved single-photon measurement setup. Here $x_2$ can be $a_2$ $(b_2)$ for Alice (Bob). The length difference between the $L$ and $S$ paths in the UI is set to cancel the time interval between the two time-bins. After the effect of Pockel cells and the unbalanced interferometer, the particle is measured in the diagonal basis in both the polarization and spatial DOF by a 50:50 BS, two PBSs oriented at $45^o$ and four single-photon detectors which are omitted in this figure.} \label{s2}
\end{figure}
In order to get a higher success probability, an improved SPM device consisting of two UIs and two Pockel cells is shown in Fig. \ref{s2}.  The length difference between the $L$ and $S$ paths is set in the same way as before. With the effect of two PCs and two UIs, the state is adjusted to
\begin{eqnarray}
  &&\vert H^SH^S\rangle_{a_1b_1}\vert V^{SL}V^{SL}\rangle_{a_{2d}b_{2d}}\nonumber\\&&+\vert V^SV^S\rangle_{a_1b_1}\vert V^{SL}V^{SL}\rangle_{a_{2u}b_{2u}}\nonumber\\&&+\vert H^LH^L\rangle_{a_1b_1}\vert H^{LS}H^{LS}\rangle_{a_{2d}b_{2u}}\nonumber\\&&+\vert V^LV^L\rangle_{a_1b_1}\vert H^{LS}H^{LS}\rangle_{a_{2u}b_{2d}}.
\end{eqnarray}
We find that both the two photons will arrive at the same time, i.e., in the middle time slot. However, there are now two potential spatial modes for each photon, the up mode ``$u$" and the down mode ``$d$". The particles are measured in the $\vert \pm\rangle$ basis in both the polarization DOF and the spatial mode. The effect of a 50:50 BS can be described as
\begin{eqnarray}
  In_u\rightarrow \frac{1}{\sqrt{2}}(Out_u+Out_d),\\
  In_d\rightarrow \frac{1}{\sqrt{2}}(Out_u-Out_d).
\end{eqnarray}
Here $In_u$ and $In_d$ denote the up and down input ports, while $Out_u$
and $Out_d$ are the two output ports of the BS. After the two particles are measured, the state of $a_1$ and $b_1$ collapses into a maximally hyperentangled state. The relationship between the measurement results and the shared states are shown in Table \ref{tab2}. The success probability using the improved measurement device is enhanced to $P_1=4|\alpha\beta\delta\eta|^2$, which is the same as that of the hyperconcentration scheme for spatial mode and polarization hyperentangled states using only linear optics \cite{hc1}.

\begin{table}[!htbp]
\caption{The relation between measurement results of $a_2b_2$ and the final state of $a_1b_1$. }\label{tab2}
\begin{tabular}{c|cl }
  \hline
   $M_{a_2b_2}$  & $\vert \Psi\rangle_{a_1b_1}$ \\ \hline\hline
  $\vert +\rangle_{a_{2u}}\vert +\rangle_{b_{2u}}$,
  $\vert -\rangle_{a_{2u}}\vert -\rangle_{b_{2u}}$ \\
  $\vert -\rangle_{a_{2d}}\vert +\rangle_{b_{2d}}$,
  $\vert +\rangle_{a_{2d}}\vert -\rangle_{b_{2d}}$
  & $\frac{1}{2}(\vert HH\rangle+\vert VV\rangle)\otimes(\vert SS\rangle+\vert LL\rangle)$&   \\ \hline
  $\vert +\rangle_{a_{2u}}\vert -\rangle_{b_{2u}}$,
  $\vert -\rangle_{a_{2u}}\vert +\rangle_{b_{2u}}$\\
  $\vert -\rangle_{a_{2d}}\vert -\rangle_{b_{2d}}$,
  $\vert +\rangle_{a_{2d}}\vert +\rangle_{b_{2d}}$
   & $\frac{1}{2}(\vert HH\rangle+\vert VV\rangle)\otimes(\vert SS\rangle-\vert LL\rangle)$ &   \\ \hline
  $\vert +\rangle_{a_{2u}}\vert -\rangle_{b_{2d}}$,
  $\vert -\rangle_{a_{2u}}\vert +\rangle_{b_{2d}}$\\
  $\vert -\rangle_{a_{2d}}\vert -\rangle_{b_{2u}}$,
  $\vert +\rangle_{a_{2d}}\vert +\rangle_{b_{2u}}$
   & $\frac{1}{2}(\vert HH\rangle-\vert VV\rangle)\otimes(\vert SS\rangle+\vert LL\rangle)$ &   \\ \hline
  $\vert +\rangle_{a_{2u}}\vert +\rangle_{b_{2d}}$,
  $\vert -\rangle_{a_{2u}}\vert -\rangle_{b_{2d}}$\\
  $\vert -\rangle_{a_{2d}}\vert +\rangle_{b_{2u}}$,
  $\vert +\rangle_{a_{2d}}\vert -\rangle_{b_{2u}}$
  & $\frac{1}{2}(\vert HH\rangle-\vert VV\rangle)\otimes(\vert SS\rangle-\vert LL\rangle)$ &   \\
  \hline
\end{tabular}
\end{table}

\begin{figure}[!h]
\centering
\subfigure[]{
\includegraphics*[width=2.5in]{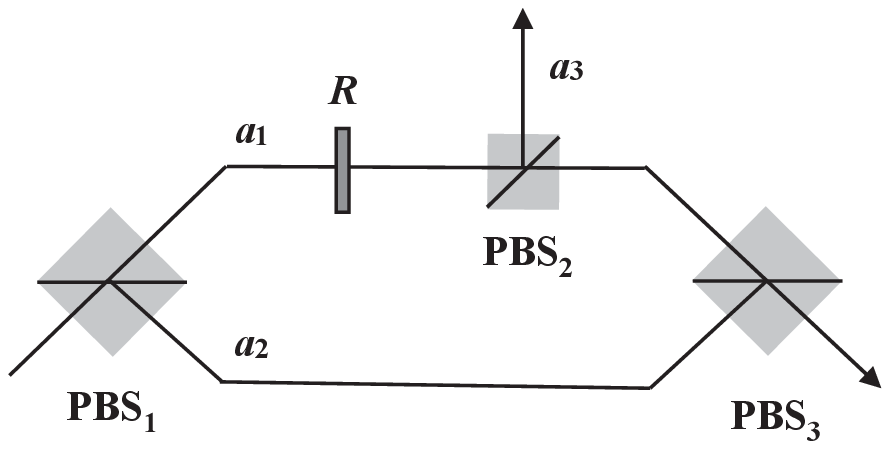} \label{c21}
}
\subfigure[]{
\includegraphics*[width=3in]{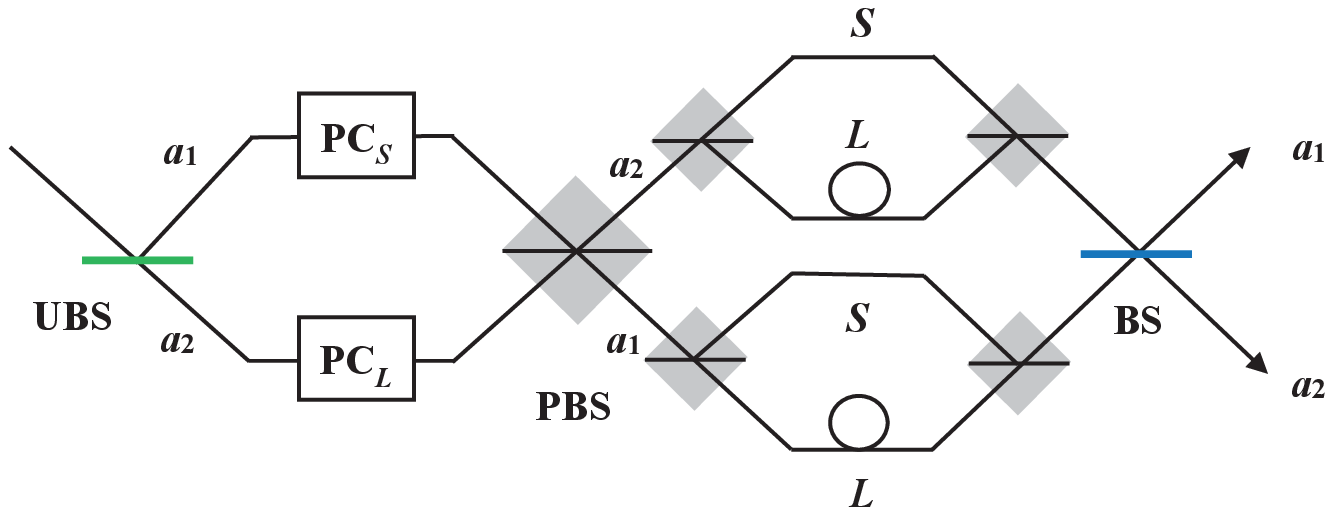} \label{c22}
}
\caption{(Color online) The schematic of our hyperentanglement concentration scheme for an initial state with known parameters. (a) This part is used to concentrate the polarization entanglement. $R$ represents a wave plate which rotates the horizontal polarization by an angle $\theta=\arccos(\beta/\alpha)$. $a_i$ $(i=1,2,3)$ is the spatial mode of particle A. The state is postselected on the condition that  the photon detector  placed in path $a_3$ (omitted in the figure) does not click, and then guided to the input port of the following device. (b) This part is used to concentrate the time-bin entanglement. UBS represents an unbalanced beam splitter with the reflection coefficient $\delta$. The desired hyperentangled state can be obtained by postselecting the situations that particle $A$ does not arrive in the middle time slot in these two potential spatial modes.}
\end{figure}

\section{Hyperentanglement concentration  with known parameters}

The schematic of our hyperentanglement concentration for a state with known parameters is shown in Fig. 4. The scheme is implemented in two steps. The first step concentrates the polarization state and the second one deals with the time-bin state. The initial state is $\vert \phi_0\rangle_{AB}$. Here we assume that $\vert \alpha\vert >\vert \beta \vert$. The entire concentration procedure can be completed by only one party, say Alice.

First, Alice guides her photon $A$ into a parameter-splitting device (Figure 4(a)) . The effect of the wave plate $R$ is
\begin{eqnarray}
  \vert H\rangle \rightarrow \cos\theta\vert H\rangle+\sin\theta\vert V\rangle
\end{eqnarray}
Where $\theta$ is adjusted to $\theta=\arccos(\beta/\alpha)$. Therefore, after passing through the wave plate, the photon state is
\begin{eqnarray}
  \vert \phi_0\rangle_{AB}&=&[\beta(\vert HH\rangle_{a_1B}+\vert VV\rangle_{a_2B})\nonumber\\&&+\sqrt{\vert \alpha\vert^2-\vert \beta\vert ^2}\vert VH\rangle_{a_1B}]\nonumber\\&&\otimes(\delta\vert SS\rangle+\eta\vert LL\rangle)
\end{eqnarray}
Here we use photon $A$'s paths $a_1$ and $a_2$ to label it. After passing through PBS$_2$ and PBS$_3$, the photon state can be written as
\begin{eqnarray}
   \vert \phi_0\rangle_{AB}&=&[\beta(\vert HH\rangle_{AB}+\vert VV\rangle_{AB})\nonumber\\&&+\sqrt{\vert \alpha\vert^2-\vert \beta\vert ^2}\vert VH\rangle_{a_3B}]\nonumber\\&&\otimes(\delta\vert SS\rangle+\eta\vert LL\rangle)
\end{eqnarray}
We can see that if particle $A$ emerges in spatial mode $a_3$, the polarization state is no longer entangled. Otherwise, a maximally entangled polarization state is obtained with probability $2\vert \beta\vert ^2$.

Then $A$ is put into the second device shown in Fig. \ref{c22}, which is used to concentrate the temporal DOF. The unbalanced BS (UBS) \cite{hc1,ubs} has a reflection coefficient $\eta$ and transmission coefficient $\delta$. Then the state evolves as
\begin{widetext}
\begin{eqnarray}
  \vert \phi_1\rangle_{AB}&=&(\delta\vert H^SH^S\rangle+ \delta\vert V^SV^S\rangle+\eta\vert H^LH^L\rangle+\eta\vert V^LV^L\rangle)_{AB}\nonumber\\
  &\xrightarrow[]{UBS}& \delta^2\vert H^S\rangle_{a_1}\vert H^S\rangle_B + \delta^2\vert V^S\rangle_{a_1}\vert V^S\rangle_B+\delta\eta \vert H^L\rangle_{a_1}\vert H^L\rangle_B+\delta\eta\vert V^L\rangle_{a_1}\vert V^L\rangle_B\nonumber\\
  &&+\delta\eta\vert H^S\rangle_{a_2}\vert H^S\rangle_B + \delta\eta\vert V^S\rangle_{a_2}\vert V^S\rangle_B+\eta^2 \vert H^L\rangle_{a_2}\vert H^L\rangle_B+\eta^2\vert V^L\rangle_{a_2}\vert V^L\rangle_B\nonumber\\
  &\xrightarrow[PC_S]{PC_L}&\delta^2\vert V^S\rangle_{a_1}\vert H^S\rangle_B + \delta^2\vert H^S\rangle_{a_1}\vert V^S\rangle_B+\delta\eta \vert H^L\rangle_{a_1}\vert H^L\rangle_B+\delta\eta\vert V^L\rangle_{a_1}\vert V^L\rangle_B\nonumber\\
  &&+\delta\eta\vert H^S\rangle_{a_2}\vert H^S\rangle_B + \delta\eta\vert V^S\rangle_{a_2}\vert V^S\rangle_B+\eta^2 \vert V^L\rangle_{a_2}\vert H^L\rangle_B+\eta^2\vert H^L\rangle_{a_2}\vert V^L\rangle_B\nonumber\\
  &\xrightarrow[]{PBS}&\delta^2\vert V^S\rangle_{a_2}\vert H^S\rangle_B + \delta^2\vert H^S\rangle_{a_1}\vert V^S\rangle_B+\delta\eta \vert H^L\rangle_{a_1}\vert H^L\rangle_B+\delta\eta\vert V^L\rangle_{a_2}\vert V^L\rangle_B\nonumber\\
  &&+\delta\eta\vert H^S\rangle_{a_2}\vert H^S\rangle_B + \delta\eta\vert V^S\rangle_{a_1}\vert V^S\rangle_B+\eta^2 \vert V^L\rangle_1\vert H^L\rangle_B+\eta^2\vert H^L\rangle_{a_2}\vert V^L\rangle_B\nonumber\\
  &\xrightarrow[]{UIs}&\delta^2\vert V^{SL}\rangle_2\vert H^S\rangle_B + \delta^2\vert H^{SL}\rangle_{a_1}\vert V^S\rangle_B+\delta\eta \vert H^{LL}\rangle_{a_1}\vert H^L\rangle_B+\delta\eta\vert V^{LL}\rangle_{a_2}\vert V^L\rangle_B\nonumber\\
  &&+\delta\eta\vert H^{SS}\rangle_{a_2}\vert H^S\rangle_B + \delta\eta\vert V^{SS}\rangle_{a_1}\vert V^S\rangle_B+\eta^2 \vert V^{LS}\rangle_{a_1}\vert H^L\rangle_B+\eta^2\vert H^{LS}\rangle_{a_2}\vert V^L\rangle_B.\nonumber\\
\end{eqnarray}
\end{widetext}
We find that by rejecting the cases that $A$ arrives in the middle time slot ($\vert SL\rangle$ and $\vert LS\rangle$), the preserved state is the desired maximally hyperentangled one. The unwanted component can be discarded by a time gate. However, the particle has two potential spatial modes. To get the desired maximally hyperentangled state, the 50:50 BS in Figure 4(b) is introduced. Then the states postselected in paths $a_1$ and $a_2$ are
\begin{eqnarray}
  \vert \Psi_{++}\rangle_{a_1B}= \frac{1}{2}(\vert HH\rangle+\vert VV\rangle)_{a_1B}\otimes(\vert S'S\rangle +\vert L'L\rangle),\nonumber\\
  \vert \Psi_{--}\rangle_{a_2B}= \frac{1}{2}(\vert HH\rangle-\vert VV\rangle)_{a_2B}\otimes(\vert S'S\rangle -\vert L'L\rangle).
\end{eqnarray}
Here we use $\vert S'\rangle$, ($\vert L'\rangle$) to represent the $\vert SS\rangle$, ($\vert LL\rangle$) time states of photon $A$. The total success probability of our hyperentanglement concentration scheme with known parameters is $P_2=4|\beta\delta\eta|^2$.

\section{Discussion and summary}
We have proposed two hyperentanglement concentration schemes for two-photon state partially hyperentangled in the time-bin and polarization DOFs. The two schemes apply to the cases where the parameters of the initial states are unknown and known, respectively. In the first scheme, two identical partially entangled states are required. Alice and Bob perform the polarization and time-bin parity check measurements, respectively. The time-bin parity check measurement is implemented using Pockel cells and polarizing beam splitters. Only when both of the two parties get the even-parity results will the selected state be the desired one.  To obtain the two-photon hyperentangled state, Alice and Bob measure two photons in the diagonal basis in both the polarization state and the time-bin DOFs. With a simple single-photon measurement device which consist of only linear optics, the success probability of the concentration is only $\vert \alpha\beta\delta\gamma\vert ^2$. We showed that this can be enhanced to $P_1=4\vert \alpha\beta\delta\gamma\vert ^2$ via an  improved measurement device. In the second scheme, only one copy of the initial state is required and only one of the two parties is needed to perform all the required local operations. The parameter splitting method is used to first concentrate the polarization DOF.
For the concentration of the time-bin state, the desired state is obtained by postselecting on the condition that the photon is not detected in the middle time slot.  The success probability is $P_2=4\vert \beta\delta\gamma\vert ^2$, where $\vert \alpha\vert >\vert \beta\vert$.

\begin{figure}[]
\centering
\includegraphics*[width=3in]{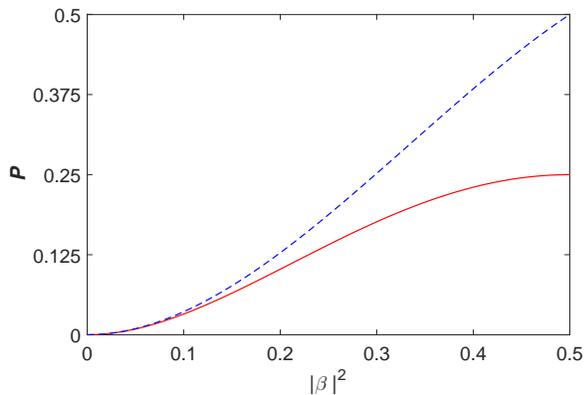}
\caption{(Color online) The success probabilities of our two hyperentanglement concentration schemes. Here we choose a kind of special state that $\vert \alpha \vert =\vert \delta\vert$ and $\vert \beta\vert =\vert \eta\vert$. The solid line and dashed line correspond to the hyperentanglement concentration with unknown and known parameters, respectively. } \label{p}
\end{figure}

In our first concentration scheme, the desired state is obtained by preserving the case where each path has one and only one photon. In a practical application, we can simply judge whether the concentration succeeds or not by the clicks in the detectors on paths $a_2$ and $b_2$. Some failing cases can be rejected by discarding the situations where there are no clicks in either of Alice or Bob's measurement devices. If each of them record a click, there are three possible scenarios - the total number of photons in modes $a_2$ and $b_2$ is 2, 3 or 4. This is because the conventional single-photon detector cannot perfectly discriminate the number of photons. Then the corresponding photon number in $a_1$ and $b_1$ is 2, 1 or 0. If these cases are mistaken as successful events, the output state is a mixed one
\begin{eqnarray}
\rho_{a_1b_1}=F_0 \rho_0+F_1 \rho_1+F_2 \rho_2.
\end{eqnarray}
Here $\rho_0=\vert vac\rangle\langle vac\vert$ denotes the vacuum state with no photons in $a_1$ and $b_1$. The probability is $F_0=\vert \alpha^2 \delta\eta \vert ^2$. $\rho_1$ represents the one-photon state in modes $a_1$ and $b_1$ with $F_1=\vert \alpha\delta\vert ^4+\vert \alpha\eta \vert^4+\vert \alpha\beta\delta^2\vert ^2+\vert \alpha\beta\eta^2\vert ^2$. $\rho_2$ corresponds to the desired one photon per path case and $F_2=P_1$. There are several ways to eliminate the vacuum and single-photon terms. First, we can replace the original detectors with some beam splitters and more detectors to detect the two photon per path cases \cite{hc1}. Second, photon-number-resolving detectors can be used to eliminate the cases with more than one photon in one spatial mode.
In our second scheme, the concentration fails if the detector in mode $a_3$ clicks. The desired state can also be obtained by postselecting the cases where the particle emits from $a_1$ or $a_2$ at the right time slots when the state is used to complete the task in quantum communication. In this case the task is accomplished although the state is destroyed.

The success probabilities of our two schemes for some special states are shown in Fig. \ref{p}. It is clear that the second method is more efficient than the first one. In general the unknown initial state parameters can be estimated by measuring a sufficient number of sample states, but this consumes extra resources. However, the second method has a higher success probability and it only requires one copy of the less-entangled state in each round of concentration. Therefore, if the number of states to be concentrated is large, the second scheme may be more efficient and practical, even if the parties must first perform state estimation. In contrast, if the number is small, the hyperentanglement concentration scheme with the Schmidt projection method may be more practical since the two parties are not required to measure sample states to estimate the parameters of the less-entangled state.

Both these two methods can be extended to concentrate the following hyperentangled $N$-photon GHZ state
\begin{eqnarray}
  \vert \phi\rangle_{AB...C}&=&(\alpha\vert HH...H\rangle +\beta\vert VV...V\rangle)_{AB...C}\nonumber\\&\otimes&(\delta \vert SS...S\rangle+\eta \vert LL...L\rangle)_{AB...C}.
\end{eqnarray}
$A,B,...C$ represent the $N$ parties who want to share one of these four maximally hyperentangled GHZ states
\begin{eqnarray}
  \vert \Psi_{\pm\pm}\rangle_{AB...C}&=&\frac{1}{2}(\vert HH...H\rangle \pm\vert VV...V\rangle)_{AB...C}\nonumber\\&\otimes&(\vert SS...S\rangle\pm\vert LL...L\rangle)_{AB...C}.
\end{eqnarray}
On one hand, when the parameters of the initial state are unknown, two identical copies of the less-entangled states are required, $\vert \phi\rangle_{A_1B_1...C_1}$ and $\vert \phi\rangle_{A_2B_2...C_2}$. First, the $N$ parties flip the polarization and time-bin states of $A_2$, $B_2$, ..., $C_2$, respectively.
Then two of the $N$ parties, say Alice and Bob perform the parity checks on $A_1,A_2$ and $B_1,B_2$, respectively and postselect the case where both of them obtain the even-parity state. Then each of the $N$ parties performs a single-photon measurement of his/her second particle.
If the $N$ parties choose the simple device shown in Fig. \ref{s1}, only the middle time slot clicks will result in the desired state, and the success probability will decrease with the growth of photon number.
 However, if all of them choose the improved SPM shown in Fig. \ref{s2}, the success probability to obtain the maximally hyperentangled state is the same as that of the two-photon hyperentanglement concentration scheme.
On the other hand, if the parameters of the  initial less-entangled $N$-photon state are known, only one copy is sufficient. One of the $N$ parties, say Alice, performs the concentration. The remaining $N-1$ parties do nothing. The $N$ parties will share the desired maximally hyperentangled state with probability $P_2=4\vert \beta\delta\eta\vert ^2$.

Most of the existing hyperentangled concentration schemes focus on states entangled in the polarization and spatial modes.  Here we have considered a different kind of hyperentanglement - that of polarization and time-bin entanglement. The success probability of our first scheme with unknown parameters achieves the same value as the protocol for polarization and spatial mode entangled states \cite{hc1,hc5} by exploiting the improved single-photon measurement scheme that we have proposed. We must admit that the success probability of our second protocol with known parameters is smaller than that of the concentration protocol for the polarization and spatial mode hyperentangled state. This is due to the challenges of working with the time-bin qubit \cite{hc1}. On the other hand, for the $N$-photon state with known parameters, we do not require auxiliary states as in Ref. \cite{hc5}, which makes our scheme easier to implement in experiments. In addition, the time-bin DOF is a stable DOF, and since it only requires one path for transmission we do not have to worry about path-length dispersion. Moreover, it saves a large amount of quantum resources in long-distance communication schemes compared to the spatial mode DOF which requires two paths for each photon's transmission. Furthermore our schemes only require linear optics which makes them experimentally feasible. All these characteristics make our schemes useful and practical, and may lead to promising applications in long-distance quantum communication in the near future.

\section*{Acknowledgement}

XL is supported by the National Natural Science
Foundation of China under Grant No.11004258 and the Fundamental Research Funds for the Central Universities under
Grant No.CQDXWL-2012-014. SG acknowledges support from the Ontario Ministry of Research and Innovation and the Natural Sciences and Engineering Research Council of Canada.


\begin{thebibliography}{99}
\bibitem{book} M. A. Nielsen and I. L. Chuang, Quantum Computation and Quantum Information (Cambridge University Press, Cambridge, 2000).
\bibitem{qkd1} A. K. Ekert, Phys. Rev. Lett. 67, 661 (1991).
\bibitem{qkd2} C. H. Bennett, G. Brassard, and N. D. Mermin, Phys. Rev. Lett. 68, 557 (1992).
\bibitem{dense1} C. H. Bennett and S. J. Wiesner, Phys. Rev. Lett. 69, 2881 (1992).
\bibitem{dense2} X. S. Liu, G. L. Long, D. M. Tong, and L. Feng, Phys. Rev. A 65, 022304 (2002).
\bibitem{tele} C. H. Bennett, G. Brassard, C. Crepeau, R. Jozsa, A. Peres, and W. K. Wootters, Phys. Rev. Lett. 70, 1895 (1993).
\bibitem{qss1} M. Hillery, V. Bu$\breve{z}$ek, and A. Berthiaume, Phys. Rev. A 59, 1829 (1999).
\bibitem{qss2} A. Karlsson, M. Koashi, and N. Imoto, Phys. Rev. A 59, 162 (1999).
\bibitem{qss3} L. Xiao, G. L. Long, F. G. Deng, and J. W. Pan, Phys. Rev. A 69, 052307 (2004).
\bibitem{qsdc1} G. L. Long and X. S. Liu, Phys. Rev. A 65, 032302 (2002).
\bibitem{qsdc2} F. G. Deng, G. L. Long, and X. S. Liu, Phys. Rev. A 68, 042317 (2003).
\bibitem{qsdc3} C. Wang, F. G. Deng, Y. S. Li, X. S. Liu, and G. L. Long, Phys. Rev. A 71, 044305 (2005).
\bibitem{type11} P. G. Kwiat, E. Waks, A. G. White, I. Appelbaum, and P. H. Eberhard, Phys. Rev. A 60, R773 (1999).
\bibitem{type12} M. Barbieri, F. De Martini, G. DiNepi, and P. Mataloni, Phys. Rev. Lett. 92, 177901 (2004).
\bibitem{preparation1} P. G. Kwiat, J. Mod. Opt. 44, 2173 (1997).
\bibitem{preparation2} T. Yang, Q. Zhang, J. Zhang, J. Yin, Z. Zhao, M. $\dot{Z}$ukowski, Phys. Rev. Lett. 95, 240406 (2005).
\bibitem{preparation3} J. T. Barreiro, N. K. Langford, N. A. Peters, and P. G. Kwiat, Phys. Rev. Lett. 95, 260501 (2005).
\bibitem{preparation4} G. Vallone, R. Ceccarelli, F. De Martini, and P. Mataloni, Phys. Rev. A 79, 030301R (2009).
\bibitem{preparation5} R. Ceccarelli, G. Vallone, F. De Martini, P. Mataloni, and A. Cabello, Phys. Rev. Lett. 103, 160401 (2009).
\bibitem{preparation6} G.Vallone, G. Donati, R. Ceccarelli, and P. Mataloni, Phys. Rev. A 81, 052301 (2010).
\bibitem{preparation7} W. B. Gao, C. Y. Lu, X. C. Yao, P. Xu, O. G$\ddot{u}$hne, A. Goebel,
Y. A. Chen, C. Z. Peng, Z. B. Chen, and J. W. Pan, Nat. Phys. 6, 331 (2010).
\bibitem{preparation8} K. Dua and C. F. Qiao, J. Mod. Opt. 59, 611 (2012).
\bibitem{dense1} J. T. Barreiro, T. C. Wei, and P. G. Kwiat, Nature Phys. 4, 282 (2008).
\bibitem{qkd} S. P. Walborn, M. P. Almeida, P. H. S. Ribeiro, and C. H. Monken, Quan. Inf. Com. 6, 336 (2006).
\bibitem{computation} B. C. Ren and F. G. Deng, Sci. Rep. 4, 4623 (2014).
\bibitem{hyper4} B. C. Ren, H. R. Wei, and F. G. Deng, Laser Phys. Lett. 10, 095202 (2013).
\bibitem{hyper5} Y. B. Sheng, F. G. Deng, and G. L. Long, Phys. Rev. A 82, 032318 (2010).
\bibitem{hyper6} B. C. Ren, H. R. Wei, M. Hua, T. Li, and F. G. Deng, Opt. Express 20, 24664 (2012).

\bibitem{depp}  Y. B. Sheng, F. G. Deng, Phys. Rev. A 81, 032307 (2010).
\bibitem{odepp} X. H. Li, Phys. Rev. A 82, 044304 (2010).
\bibitem{odepp1} Y. B. Sheng, F. G. Deng, Phys. Rev. A 82, 044305 (2010).
\bibitem{omdepp} F. G. Deng, Phys. Rev. A 83 062316 (2011).
\bibitem{puri pdc} C. Simon and J. W. Pan, Phys. Rev. Lett. 89, 257901 (2002).
\bibitem{puri1}  Y. B. Sheng, F. G. Deng, and H. Y. Zhou, Phys. Rev. A 77 062325 (2008).
\bibitem{puri2} C. H. Bennett, G. Brassard, S. Popescu, B. Schumacher, J. A.
Smolin, and W. K. Wootters, Phys. Rev. Lett. 76, 722 (1996).
\bibitem{bsa1} P. G. Kwiat and H. Weinfurter, Phys. Rev. A 58, R2623 (1998).
\bibitem{bsa2} S. P. Walborn, S. P$\acute{a}$dua, and C. H. Monken, Phys. Rev. A 68,042313 (2003).
\bibitem{bsa3} C. Schuck, G. Huber, C. Kurtsiefer, and H. Weinfurter, Phys. Rev. Lett. 96, 190501 (2006).
\bibitem{bsa4} M. Barbieri, G. Vallone, P. Mataloni, and F. De Martini, Phys. Rev. A 75, 042317 (2007).
\bibitem{gsa} S. Y. Song, Y. Cao, Y. B. Sheng, and G. L. Long, Quan. Inf. Proc. 12, 381 (2013).
\bibitem{concen1} C. H. Bennett, H. J. Bernstein, S. Popescu, and B. Schumacher, Phys. Rev. A 53, 2046 (1996).
\bibitem{concen swap1} S. Bose, V. Vedral, and P. L. Knight, Phys. Rev. A 60, 194 (1999).
\bibitem{concen swap2} B. S. Shi, Y. K. Jiang, and G. C. Guo, Phys. Rev. A 62, 054301 (2000).
\bibitem{concen pbs1} T. Yamamoto, M. Koashi, and N. Imoto, Phys. Rev. A 64, 012304 (2001).
\bibitem{concen pbs2} Z. Zhao, J. W. Pan, and M. S. Zhan, Phys. Rev. A 64, 014301 (2001).
\bibitem{concen sheng1} Y. B. Sheng, F. G. Deng, and H. Y. Zhou, Phys. Rev. A 77, 062325 (2008).
\bibitem{concen sheng4} Y. B. Sheng, L. Zhou, S. M. Zhao, and B. Y. Zheng, Phys. Rev. A 85, 012307 (2012).
\bibitem{concen deng} F. G. Deng, Phys. Rev. A 85, 022311 (2012).
\bibitem{hc1} B. C. Ren, F. F. Du, and F. G. Deng, Phys. Rev. A 88, 012302 (2013).
\bibitem{hc5} X. H. Li, S. Ghose, Laser Phys. Lett. 11, 125201 (2014).
\bibitem{hc2} B. C. Ren and F. G. Deng, Laser Phys. Lett. 10, 115201 (2013).
\bibitem{hp} B. C. Ren, F.F. Du, and F.G. Deng, Phys. Rev. A 90, 052309 (2014).
\bibitem{hc3} B. C. Ren and G. L. Long, Opt. Express 22, 6547 (2014).

\bibitem{hc4} X. H. Li, X. Chen, and Z. Zeng, J. Opt. Soc. Am. B 30, 2774 (2013).
\bibitem{hcc} X. Chen, Z. Zeng, X. H. Li, Commun. Theor. Phys. 61, 322 (2014).

\bibitem{time} J. Brendel, N. Gisin, W. Tittel, and H. Zbinden, Phys. Rev. Lett. 82, 2594 (1999).

\bibitem{sw} Y. Soudagar, F. Bussi\`{e}res, G. Berl\'{\i}n, S. Lacroix, J. M. Fernandez, and N. Godbout, J. Opt. Soc. Am. B 24, 226 (2007).
\bibitem{PC} D. Kalamidas, Phys. Lett. A 343, 331 (2005).
\bibitem{ubs} M. Reck, A. Zeilinger, H. J. Bernstein, and P. Bertani, Phys. Rev. Lett. 73, 58 (1994).

\end{thebibliography}
\end{document}